\documentclass[aps,prl,tightenlines,twocolumn,superscriptaddress,showpacs]{revtex4}
\usepackage{amssymb}
\usepackage{color,graphicx}
\usepackage{lmodern}
\usepackage{amsmath}
\usepackage{amsbsy}
\usepackage{amsthm}
\usepackage{float}
\usepackage{bbm}
\usepackage{bm}
\usepackage{epsfig}

\begin{document}

\title{Dissipative Vibrational Model for Chiral Recognition in Olfaction}

\author{Arash Tirandaz}
\email[]{arash85201@ipm.ir}
\affiliation{Foundations of Physics Group, School of Physics, Institute for Research in Fundamental Sciences (IPM), P.O. Box 19395-5531, Tehran, Iran}

\author{Farhad Taher Ghahramani}
\email[]{farhadtqm@ipm.ir}
\affiliation{Foundations of Physics Group, School of Physics, Institute for Research in Fundamental Sciences (IPM), P.O. Box 19395-5531, Tehran, Iran}

\author{Afshin Shafiee}
\email[]{shafiee@sharif.edu}
\affiliation{Foundations of Physics Group, School of Physics, Institute for Research in Fundamental Sciences (IPM), P.O. Box 19395-5531, Tehran, Iran}
\affiliation{Research Group On Foundations of Quantum Theory and Information, Department of Chemistry, Sharif University of Technology, P.O.Box 11365-9516, Tehran, Iran}

\begin{abstract}
  We examine the olfactory discrimination of left- and right-handed enantiomers of chiral odorants based on the odorant-mediated electron transport from a donor to an acceptor of the olfactory receptors embodied in a biological environment. The chiral odorant is effectively described by an asymmetric double-well potential whose minima are associated to the left- and right-handed enantiomers. The introduced asymmetry is considered as an overall measure of chiral interactions. The biological environment is conveniently modeled as a bath of harmonic oscillators. The resulting Spin-Boson model is adapted by a polaron transformation to derive the corresponding Born-Markov master equation with which we obtain the elastic and inelastic electron tunneling rates. We show that the inelastic tunneling through left- and right-handed enantiomers occurs with different rates. The discrimination mechanism depends on the ratio of tunneling frequency to localization frequency.
\end{abstract}


\maketitle

\section{I. Introduction}
Olfaction, the sense of smell, is one of the most ancient yet intriguing senses of living organisms to being in contact with the environment. Smell is caused by the small, neutral, volatile molecules known as odorant. In Human beings, olfaction occurs when the odorant molecules bind to specific sites on olfactory receptors in nasal cavity. In spite of considerable progress towards the understanding of structure of olfactory receptors that involved in the early stages of olfactory process, the detailed mechanisms of discrimination between vast number of odorants are not yet fully understood~\cite{Zar}. The most obvious characteristic of an odorant molecule is its shape. Accordingly, Amoore first conjectured that the response to scent is initiated by a mutual structural fit between the receptor and the odorant ({\it lock and key} model)~\cite{Amo}. A more flexible modification of this idea is that the whole system distorts to induce a more appropriate mutual fit ({\it hand and glove} model). A further modification of this shape-based theory requires that a particular receptor responds to only one structural feature, such as a functional group, as opposed to the main body of the odorant ({\it odotope} model)~\cite{Mor}. In spite of the predictive power of these shape-based models, there are some evidence against them: some odorants smell the same although they are structurally very different~\cite{Tur}. Furthermore, some odorants with almost identical shapes smell very different~\cite{Ben,Bro}. The only contender to the shape-based models is the vibration-based model: an unique scent is attributed to the unique vibrational spectrum of the odorant~\cite{Dys,Wri}. Motivated by this idea, Turin proposed that mechanism of olfactory recognition is an odorant-mediated inelastic electron tunneling (ET) at the receptor: signal transduction is based on the success of an electron tunneling from a donor (D) state of a receptor to an acceptor (A) state of the same or another receptor, facilitated by a vibrational transition in the odorant corresponding to the energy difference between these states~\cite{Tur2}. Recently, Brookes and co-workers expanded this idea, and advanced a semi-classical model to show that such a mechanism fits the observed features of smell~\cite{Bro2}. They found that the rate of odorant-mediated inelastic ET is larger than the rate of elastic one. Following this approach, in a very recent work, Ch\c{e}ci\'{n}ska and co-workers examined dynamically dissipative role of environment in vibration-based olfactory recognition and showed that the strong coupling to the environment can enhance the odorant frequency resolution in the ET rates~\cite{Che}.\\
\indent The vibration-based theories of olfaction are commonly refused on the grounds that enantiomers of chiral odorants have the same vibrational spectra but different smells~\cite{Ben}. The advocates of such theories suggest that other molecular features such as structural flexibility should be included in the model to account this seemingly counterintuitive behaviour~\cite{Bro3}. To address this problem, we propose that the chiral recognition in olfaction is reliant on detection of energy difference between two enantiomers of the chiral odorant. This energy difference is resulted from chiral interactions between the odorant and the receptor. At sufficiently low temperatures, two enantiomers of a chiral molecule can be effectively described as two localized states of an asymmetric double-well potential. This asymmetry portrays the energy difference between two enantiomers. The biological environment of olfactory system can be represented as a collection of harmonic oscillators. When the oscillators couple linearly to the double-well, the result is the Spin-Boson model studied extensively in the literature, particularly by Leggett and co-workers~\cite{Leg}. Here, we examine the dynamics of olfactory ET using the Spin-Boson model and thereby obtain the elastic and inelastic ET rates for each enantiomers of a typical chiral odorant. We show that the inelastic ET rates can be different for two enantiomers. \\
\indent The article is organized as follows. In section II, we describe the olfactory Spin-Boson model for a chiral odorant. In section III, after an appropriate polaron transformation, we characterize the dynamics of the model using the Born-Markov master equation. In section IV, we conduct the resulting equations for the appropriate choice of model parameters to obtain the corresponding (in)elastic tunneling rates. Finally, we summarize our findings in the last section. Throughout this article, we suppose $\hbar=1$.
\section{II. Chiral Olfaction Model}
The free Hamiltonian of the total system consisting of the electron at the receptor, the chiral odorant and the surrounding environment can be written as
\begin{equation}\label{1}
\hat H_{\mbox{\tiny${\mbox{\tiny$\circ$}}$}}=\hat H_{e}+\hat H_{o}+\hat H_{E}
\end{equation}
The electron state at donor and acceptor sites of receptor are represented by $|D\rangle$ with energy $\varepsilon_{D}$ and $|A\rangle$ with energy $\varepsilon_{A}$, respectively. We assume that $\varepsilon_{D}>\varepsilon_{A}$, so that the electron tunneling corresponds to a vibrational absorption in the odorant. So, we can describe the electron at the receptor with Hamiltonian
\begin{equation}\label{2}
\hat H_{e}=\varepsilon_{D}|D\rangle\langle D|+\varepsilon_{A}|A\rangle\langle A|
\end{equation}
A chiral odorant can occur at least as two chiral enantiomers through the inversion at molecule's center of mass by a long-amplitude vibration known as contortional vibration~\cite{Tow,Her}. This vibration can be effectively described by an asymmetric double-well potential. The minima, associated to two chiral states $|L\rangle$ and $|R\rangle$, are separated by barrier $V_{\mbox{\tiny${\mbox{\tiny$\circ$}}$}}$. In the limit ${{V}_{\mbox{\tiny${\mbox{\tiny$\circ$}}$}}}\gg {\omega}_{\mbox{\tiny${\mbox{\tiny$\circ$}}$}}\gg k_{\mbox{\tiny$B$}}T$ (${\omega}_{\mbox{\tiny${\mbox{\tiny$\circ$}}$}}$ is the vibration frequency in each well), state space of the molecule are effectively confined in two-dimensional Hilbert space spanned by two chiral states. For most chiral molecules this limit holds up to room temperature~\cite{Tow,Her}. The left- and right-handed enantiomers of the chiral odorant are associated to states $|L\rangle$ and $|R\rangle$ localized at left and right wells of the potential, respectively. If the barrier is high enough to prevent the tunneling process, the molecule remains in its initial chiral state. The corresponding Hamiltonian of the molecule in the chiral basis is given by~\cite{Wei}
\begin{equation}\label{3}
\hat H_{o}=-\frac{\delta}{2}\hat\sigma_{x}-\frac{\omega_{z}}{2}\hat\sigma_{z}
\end{equation}
where $\delta$ is the frequency of tunneling between two chiral states and $\omega_{z}$, known as localization frequency, is the energy difference between two chiral states. The localization frequency is an overall measure of chiral interactions. The odorant's states of energy are described by a superposition of chiral states as
\begin{align}\label{4}
|1\rangle&=\sin(\frac{\theta}{2})|L\rangle+\cos(\frac{\theta}{2})|R\rangle \nonumber \\
|2\rangle&=\cos(\frac{\theta}{2})|L\rangle-\sin(\frac{\theta}{2})|R\rangle
\end{align}
where we defined $\theta=\arctan(\delta/\omega_{z})$. The energies corresponding to these states are $\mp\sqrt{\delta^{2}+\omega_{z}^{2}}$. \\
\indent The biological environment is a kind of condensed bath, conveniently modelled as a collection of harmonic oscillators with Hamiltonian
\begin{equation}\label{5}
\hat H_{E}=\sum_{i}\omega_{i}\hat b_{i}^{\dagger}\hat b_{i}
\end{equation}
where $\hat b_{i}^{\dagger}$ and $\hat b_{i}$ are the creation and annihilation operators for modes of frequency $\omega_{i}$ in the environment.\\
\indent The interaction Hamiltonian has three contributions: between donor and acceptor of the receptor with tunneling strength $\Delta$, between the donor (acceptor) and the odorant with coupling strength $\gamma_{D}$ ($\gamma_{A}$), and between the donor (acceptor) and $i$-th harmonic oscillator of the environment with coupling strength $\gamma_{iD}$ ($\gamma_{iA}$). So, the interaction Hamiltonian of total system is given by
\begin{align}\label{6}
\hat H_{int}&=\Delta(|A\rangle\langle D|+|D\rangle\langle A|)\nonumber \\ &+(\gamma_{D}|D\rangle\langle D|+\gamma_{A}|A\rangle\langle A|)\hat\sigma_{x}\nonumber \\ &+\sum_{i}(\gamma_{iD}|D\rangle\langle D|+\gamma_{iA}|A\rangle\langle A|)(\hat b_{i}^{\dagger}+\hat b_{i})
\end{align}
\section{III. Dynamics}
\begin{figure}[H]
  \centering
  \includegraphics[width=0.46\textwidth]{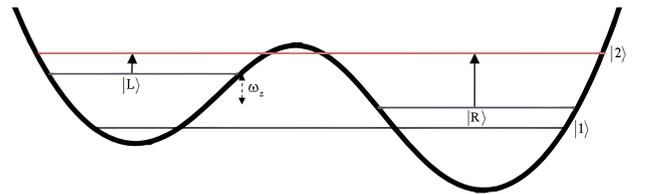}
  \caption{Relevant transitions in chiral odorant.}
  \label{Fig1}
\end{figure}

In order to avoid the undesired transitions from donor to acceptor when the odorant is absent, the coupling strength $\Delta$ should be small compared to other energy scales in the model. In this regime, we conveniently move into a polaron transformed reference frame. We define the unitary polaron transformation as
\begin{align}\label{7}
 U&=\exp\Big[\frac{i}{2}(\lambda'|D\rangle\langle D|-\lambda|A\rangle\langle A|)\hat\sigma_{y}\Big]\nonumber \\
 &\times\exp\Big[\sum_{i}(\frac{\gamma_{iD}}{\omega_{i}}|D\rangle\langle D|+\frac{\gamma_{iA}}{\omega_{i}}|A\rangle\langle A|)(\hat b_{i}^{\dagger}-\hat b_{i})\Big]
\end{align}
where we defined
\begin{align}\label{8}
  \lambda&=\tan^{-1}\big(\frac{\delta-\gamma_{D}}{\omega_{z}}\big)\nonumber \\
  \lambda'&=\tan^{-1}\big(\frac{\delta+\gamma_{A}}{\omega_{z}}\big)
\end{align}
Under the polaron transformation, the free Hamiltonian and interaction Hamiltonian take the form
\begin{equation}\label{9}
\widetilde{H}_{\mbox{\tiny${\mbox{\tiny$\circ$}}$}}\!=\!(\varepsilon_{D}\!+\!\mu'\hat\sigma_{z})|D\rangle\!\langle D|+(\varepsilon_{A}\!+\!\mu\hat\sigma_{z})|A\rangle\!\langle A|
+\!\sum_{i}\!\omega_{i}\hat b_{i}^{\dagger}\hat b_{i}
\end{equation}
with
\begin{align}\label{10}
\mu&=-\frac{\omega_{z}}{2}\cos{\lambda}\Big[1+\Big(\frac{\delta-\gamma_{D}}{\omega_{z}}\Big)^{2}\Big]\nonumber \\
\mu'&=-\frac{\omega_{z}}{2}\cos{\lambda'}\Big[1+\Big(\frac{\delta+\gamma_{A}}{\omega_{z}}\Big)^{2}\Big]
\end{align}
and
\begin{align}\label{11}
 \widetilde{H}_{int}=&\Delta\Big(|A\rangle\langle D|e^{\frac{i}{2}(\lambda+\lambda')\hat\sigma_{y}}\hat\digamma_{+}  \nonumber \\
 & +|D\rangle\langle A|e^{-\frac{i}{2}(\lambda+\lambda')\hat\sigma_{y}}\hat\digamma_{-}\Big)
\end{align}
with
\begin{equation}\label{12}
\hat\digamma_{\pm}=e^{\pm\sum_{i}(\frac{\gamma_{iD}-\gamma_{iA}}{\omega_{i}})(\hat b_{i}^{\dagger}-\hat b_{i})}
\end{equation}
The dynamics of the reduced density matrix of the receptor and the odorant (hereafter receptor+odorant) can be described by the so-called Redfield equation~\cite{Sch}
\begin{equation}\label{13}
\partial_{t}\hat\rho(t)=-\int_{0}^{t}dt'~Tr_{E}\big[\widetilde{H}_{int}(t),[\widetilde{H}_{int}(t'),\hat\rho_{tot}(t')]\big]
\end{equation}
If we assume that the interaction between the receptor+odorant and the environment is sufficiently weak, the total density matrix remains at all times in an approximate product form ($\hat\rho_{tot}(t)\approx\hat\rho(t)\otimes\hat\rho_{E}(t)$). Also, because the environment is large in comparison with the size of the receptor+odorant, the temporal change of the environment density matrix can be neglected (Born approximation) ($\hat\rho_{E}(t)\approx\hat\rho_{E}(0))$. At sufficiently high temperatures, moreover, it is assumed that the environment quickly forget any internal self-correlations established in the course of the interaction with the receptor+odorant (Markov approximation). After some mathematics, then the Born-Markov master equation is obtained as
\begin{equation}\label{14}
\partial_{t}\hat\rho(t)=-\imath\big[\hat H^{RO}_{\mbox{\tiny${\mbox{\tiny$\circ$}}$}},\hat\rho(t)\big]-\pounds_{\rho}(t)
\end{equation}
with
\begin{align}\label{15}
  \pounds_{\rho}(t)&=\int_{0}^{\infty} d\tau \Big\{C(\tau)\big[\hat H^{RO}_{int},\hat H^{RO}_{int}(-\tau)\hat\rho(t)\big] \nonumber \\ & \quad \quad \quad \quad +C(-\tau)\big[\hat\rho(t)\hat H^{RO}_{int}(-\tau),\hat H^{RO}_{int}\big]\Big\}
\end{align}
where
\begin{equation}\label{16}
\hat H^{RO}_{int}(-\tau)=e^{it\varepsilon}|A\rangle\langle D|\hat\Xi(\tau)+h.c
\end{equation}
in which $\varepsilon=\varepsilon_{D}-\varepsilon_{A}$ and we defined
\begin{equation}\label{17}
\hat\Xi(\tau)=\left(\begin{array}{cc}
\Xi_{1}(\tau) & \Xi_{2}(\tau) \\
\Xi_{3}(\tau) & \Xi_{4}(\tau)
\end{array} \right)
\end{equation}
where
\begin{align}\label{18}
 \Xi_{1}(\tau)&=\cos(\frac{\lambda+\lambda'}{2})e^{-i\tau(\mu-\mu')} \nonumber \\
 \Xi_{2}(\tau)&=\sin(\frac{\lambda+\lambda'}{2})e^{-i\tau(\mu+\mu')} \nonumber \\
 \Xi_{3}(\tau)&=-\sin(\frac{\lambda+\lambda'}{2})e^{i\tau(\mu+\mu')} \nonumber \\
 \Xi_{4}(\tau)&=\cos(\frac{\lambda+\lambda'}{2})e^{i\tau(\mu-\mu')}
\end{align}
and the self-correlation functions of the environment $C(\tau)$ are obtained by
\begin{align}\label{19}
 C(\tau)&=C^{\ast}(-\tau)=tr_{E}\Big\{\hat\digamma_{\pm}(-\tau)\hat\digamma_{\mp}(0)\Big\}\nonumber \\
 &=e^{-\big\{\nu(\tau)+i\eta(\tau)\big\}}
\end{align}
where $\nu(\tau)$ and $\eta(\tau)$ are the noise and dissipation kernels
\begin{align}\label{20}
  \nu(\tau)&=\int_{0}^{\infty}d\omega\frac{J(\omega)}{\omega^{2}}\Big[\sin^{2}(\frac{\omega\tau}{2})\coth(\frac{\omega}{2k_{\mbox{\tiny$B$}}T})\Big] \nonumber \\
  \eta(\tau)&=\int_{0}^{\infty}d\omega\frac{J(\omega)}{\omega^{2}}\sin(\omega\tau)
\end{align}
Here, $J(\omega)$ is the spectral density of a continuous spectrum of environmental frequencies, $\omega$. It encapsulates the physical properties of the environment. For more convenience, we employ an ohmic spectral density with an exponential cut-off as
\begin{equation}\label{21}
J(\omega)=\sum_{i}(\gamma_{iD}-\gamma_{iA})^{2}\delta(\omega-\omega_{i})\equiv J_{\mbox{\tiny${\mbox{\tiny$\circ$}}$}}\omega e^{-\omega/\Lambda}
\end{equation}
in which $J_{\mbox{\tiny${\mbox{\tiny$\circ$}}$}}$ is a measure of system-environment coupling strength and $\Lambda$ is a high-frequency cut-off. If we replace the explicit form of spectral density into (\ref{20}), and for the high-temperature environment, approximate the term $\coth(\omega/2k_{\mbox{\tiny$B$}}T)$ by $2k_{\mbox{\tiny$B$}}T/\omega$, we obtain
\begin{equation}\label{22}
C(\tau)=e^{-\frac{k_{\mbox{\tiny$B$}}TJ_{\mbox{\tiny${\mbox{\tiny$\circ$}}$}}}{\Lambda}
\!\big[\Lambda\tau\tan^{-1}(\Lambda\tau)-\frac{\ln(1+\Lambda^{2}\tau^{2})}{2}\big]\!+i\tan^{-1}(\Lambda\tau)}
\end{equation}
If we expand the the exponential function to second order in $\tau$ we arrive at
\begin{equation}\label{23}
C(\tau)\approx e^{-k_{B}TJ_{\mbox{\tiny${\mbox{\tiny$\circ$}}$}}\Lambda\tau^{2}-i\Lambda\tau}
\end{equation}
\noindent The elastic ET coincides with a situation in which the chiral odorant is present in the receptor but does not undergo any transition. The inelastic ET through left- and right-handed enantiomers are accompanied by a vibrational transition from left- and right-handed states to the first excited energy state, respectively (FIG.~\ref{Fig1}).
\noindent Now we can derive the explicit expressions of the corresponding elastic and inelastic ET rates. We may decompose the receptor+odorant density matrix as $\hat\rho(t)=\sum_{i,j;X,X}\rho_{ijXX'}|i,X\rangle\langle j,X'|$ where $i,j\in L,R$ and $X,X'\in D,A$. In accordance with previous works, we assume that the donor state of the receptor only couples to the lower state of the odorant at all times. In the decoherence regime where the unitary dynamics is ignored, the population dynamics of acceptor state $|A\rangle$ and odorant state $|j\rangle$ is given by $\langle A,j|\pounds_{\rho}(t)|A,j\rangle$. We may then obtain the elastic and inelastic ET rate for transition $|D,i\rangle\rightarrow|A,f\rangle$ by
\begin{align}\label{24}
\Gamma_{DiAj}&=2\Delta^{2}\Re\int_{0}^{\infty} d\tau\Big\{C(\tau)e^{i\varepsilon\tau}\langle f|\hat\Xi(\tau)|i\rangle\nonumber \\ &\qquad\qquad\qquad\qquad\langle i|e^{-\frac{i}{2}(\lambda+\lambda')\hat\sigma_{y}}|j\rangle\Big\}
\end{align}
Using the explicit form of the environmental correlation function $C(\tau)$ and the transition matrix $\Xi(\tau)$, we finally obtain the inelastic ET rates as
\begin{align}\label{25}
\Gamma_{DLA2}&\approx\cos\big(\frac{\lambda+\lambda'-\theta}{2}\big)\sin\big(\frac{\lambda+\lambda'}{2}\big)\sin\big(\frac{\theta}{2}\big)  \nonumber \\ &\qquad \frac{\sqrt{\pi}\Delta^{2}}{\sqrt{k_{B}TJ_{\mbox{\tiny${\mbox{\tiny$\circ$}}$}}\Lambda}}
\exp\Big\{{\frac{-[\varepsilon+(\mu+\mu')-\Lambda]}{4k_{B}TJ_{\mbox{\tiny${\mbox{\tiny$\circ$}}$}}\Lambda}}\Big\}\nonumber \\
\Gamma_{DRA2}&\approx\sin\big(\frac{\lambda+\lambda'-\theta}{2}\big)\sin\big(\frac{\lambda+\lambda'}{2}\big)\cos\big(\frac{\theta}{2}\big)  \nonumber \\ &\qquad\frac{\sqrt{\pi}\Delta^{2}}{\sqrt{k_{B}TJ_{\mbox{\tiny${\mbox{\tiny$\circ$}}$}}\Lambda}}
\exp\Big\{{\frac{-[\varepsilon-(\mu+\mu')-\Lambda]}{4k_{B}TJ_{\mbox{\tiny${\mbox{\tiny$\circ$}}$}}\Lambda}}\Big\}
\end{align}
The elastic ET rate $\Gamma_{DiAi}$ ($i=L,R$), which is equal for two enantiomers, is given by
\begin{align}\label{26}
\Gamma_{DiAi}&=\cos^{2}\big(\frac{\lambda+\lambda'}{2}\big) \nonumber \\
  &\qquad \frac{\sqrt{\pi}\Delta^{2}}{\sqrt{k_{B}TJ_{\mbox{\tiny${\mbox{\tiny$\circ$}}$}}\Lambda}}
\exp\Big\{{\frac{-(\varepsilon-\Lambda)}{4k_{B}TJ_{\mbox{\tiny${\mbox{\tiny$\circ$}}$}}\Lambda}}\Big\}
\end{align}
In the limit $\delta\rightarrow0$, our expressions for the (in)elastic ET rates reduce to the corresponding expressions for an odorant with one configuration obtained by Ch\c{e}ci\'{n}ska and co-workers~\cite{Che}.
\section{IV. Results}
\begin{figure}[H]
  \centering
  \includegraphics[width=0.46\textwidth]{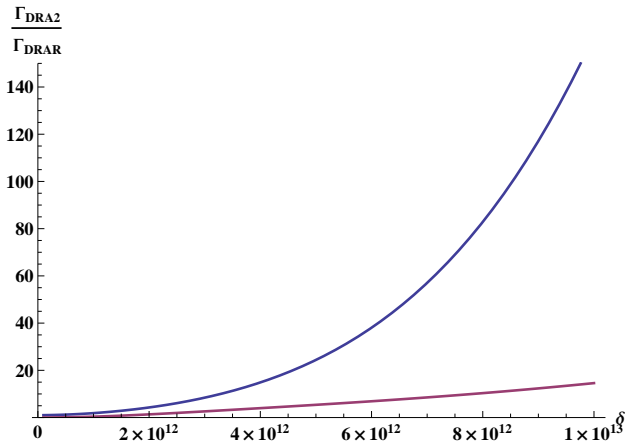}
  \caption{The ratio of inelastic rate to elastic rate versus the tunneling frequency $\delta$ for left-handed enantiomer (blue) and right-handed enantiomer (red) at $\omega_{z}=10^{12}Hz$.}
  \label{Fig2}
\end{figure}

\begin{figure}[H]
  \centering
  \includegraphics[width=0.5\textwidth]{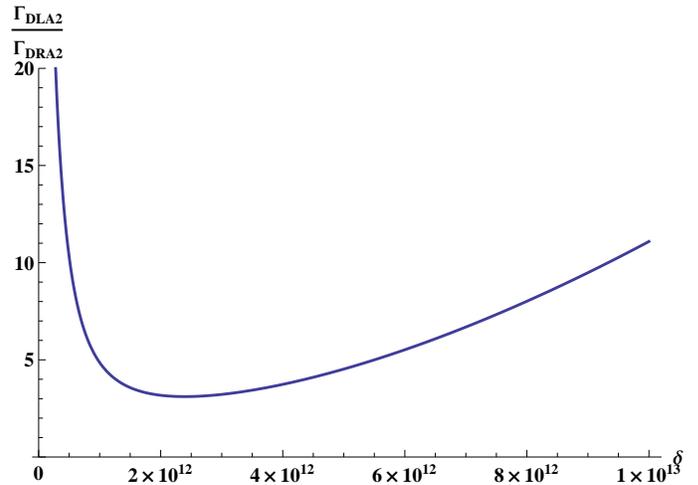}
  \caption{The ratio of inelastic rates for the left-handed enantiomer $\Gamma_{DLA2}$ to the right-handed enantiomer$\Gamma_{DRA2}$ versus the tunneling frequency $\delta$ at $\omega_{z}=10^{12}Hz$.}
  \label{Fig3}
\end{figure}

To calculate the numerical values of corresponding (in)elastic ET rates, we first estimate the parameters relevant to our analysis. The magnitude of odorant's tunneling frequency, $\delta$, extracted from the spectroscopic data, ranges from the inverse of the lifetime of the universe to millions of hertz~\cite{Qua}. We consider this parameter as the signature of the chiral odorant. The localization frequency of the odorant, $\omega_{z}$, represents an overall measure of all chiral interactions involved. A chiral interaction in general transforms as a pseudo-scalar~\cite{Bar}. For our system, these chiral interactions include the dispersion intermolecular interactions between the odorant and the receptor. Since the structural characteristics of odorants and especially receptors are complex by nature, to our knowledge, the molecular interactions between them are not characterized yet. Since both odorant and receptor are chiral, nevertheless, we assume that the contribution of intermolecular chiral interactions is significant. Note that if chiral interactions are strong enough to overcome tunneling process, say $\omega_{z}\gg\delta$, they confine the molecule in the chiral state corresponding to the deeper well (the Hamiltonian (\ref{3}) coincide only with the right-handed enantiomer). This situation may be realized in chiral biomolecules with only one stable enantiomer, e.g., the building blocks of life, i.e., L-proteins and D-sugars. For chiral odorants with two stable enantiomers, we exclude this situation. On the other hand, as the discrimination factor of our model is the localization frequency, $\omega_{z}$, in the limit $\delta\gg\omega_{z}$, we obtain the same inelastic ET rates for the left- and right-handed enantiomers. Accordingly, we consider the case in which tunneling and localization compete with each other. We assume that the DA energy gap, $\varepsilon$, is relatively close to resonance with corresponding transitions in the odorant. The energy difference between left-handed (right-handed) enantiomer and the first excited state is $\sqrt{\delta^{2}+\omega_{z}^{2}}-\omega_{z}/2$ ($\sqrt{\delta^{2}+\omega_{z}^{2}}+\omega_{z}/2$). So, we choose the mean value for the DA energy gap, say $\varepsilon=\sqrt{\delta^{2}+\omega_{z}^{2}}$. Also, we take this value as the characteristic frequency of the odorant. While the odorant's structure is known, the detailed structure of donor and acceptor sites of the receptor interacting with it is relatively unknown. Since the elastic ET increases with the DA tunneling frequency, $\Delta$, which is undesired as far as the odorant-assisted mechanism is concerned, we keep $\Delta$ small in comparison to the characteristic frequency of the odorant. So, we estimate $\Delta\simeq0.01\sqrt{\delta^{2}+\omega_{z}^{2}}$. The estimation of coupling frequency between the DA pair and the odorant mode requires more consideration. The coupling strength between donor (acceptor) $\gamma_{D}$ ($\gamma_{A}$) is proportional to change of the odorant's intrinsic dipole moment due to the interaction with the electric field of the transferred electron localized at the donor (acceptor) site of the receptor. If we suppose that the force on the odorant molecule as a result of electron's electric field at the donor site is opposite of the of the corresponding force at the acceptor site, we have $\gamma_{D}=-\gamma_{A}$ (see~\cite{Bro2}). On the other hand, the difference between DA-odorant couplings $\gamma_{D}-\gamma_{A}$, calculated from the associated Huang-Rhys factor, transcribed to our system, is given as $0.1\sqrt{\delta^{2}+\omega_{z}^{2}}$~\cite{Bro2}. So, we estimate $\gamma_{D}=-\gamma_{A}=0.05\sqrt{\delta^{2}+\omega_{z}^{2}}$.
The spectral density is obtained from the microscopic details of the model under consideration. The simplest model arises when the odorant is treated as a point dipole inside a uniform, spherical protein surrounded by a uniform polar solvent. For a Debye solvent and a protein with a static dielectric constant, the parameters of the spectral density (\ref{20}) is obtained as~\cite{Gil}
\begin{equation}\label{27}
J_{\mbox{\tiny${\mbox{\tiny$\circ$}}$}}=\frac{(\Delta\nu)^{2}}{4\pi\epsilon_{\mbox{\tiny${\mbox{\tiny$\circ$}}$}}b^{3}}\frac{6\epsilon_{p}(\epsilon_{s}-\epsilon_{\infty})}{(2\epsilon_{s}+\epsilon_{p})(2\epsilon_{\infty}+\epsilon_{p})\Lambda}
\end{equation}
where
\begin{equation}
\label{28}
\Lambda=\frac{2\epsilon_{s}+\epsilon_{p}}{2\epsilon_{\infty}+\epsilon_{p}}\tau^{-1}_{D}
\end{equation}
in which $\Delta\nu$ is the difference between the dipole moment of the odorant in the ground and excited states, $b$ is the radius of the protein containing the odorant, $\epsilon_{p}$ is the dielectric constant of the protein environment, $\epsilon_{s}$ and $\epsilon_{\infty}$ are the static and high-frequency dielectric constants of the solvent, respectively and $\tau_{D}$ is the Debye relaxation time of the solvent. For an odorant in water, we have $J_{\mbox{\tiny${\mbox{\tiny$\circ$}}$}}\approx1$ and $\Lambda\approx10^{12}Hz$. Accordingly, we summarize the relevant parameters in TABLE~I.
\begin{table}[H]
\begin{center}
\def\arraystretch{1.2}
\begin{tabular} {c c c c}
  Parameter &\hspace{0.4cm} Value &\hspace{0.4cm} Parameter &\hspace{0.4cm} Value \\
  \hline
  $\varepsilon$ &\hspace{0.4cm} $\sqrt{\delta^{2}+\omega_{z}^{2}}$ &\hspace{0.4cm} $J_{\mbox{\tiny${\mbox{\tiny$\circ$}}$}}$ &\hspace{0.4cm}  $1$ \\
  $\Delta$ &\hspace{0.4cm} $0.01\sqrt{\delta^{2}+\omega_{z}^{2}}$ &\hspace{0.4cm} $\Lambda$&\hspace{0.4cm}  $10^{12}Hz$ \\
  $\gamma_{D}(=-\gamma_{A})$ &\hspace{0.4cm} $0.1\sqrt{\delta^{2}+\omega_{z}^{2}}$ &\hspace{0.4cm} $T$ &\hspace{0.4cm}  300K \\
\end{tabular}
\caption{Parameter values for the chiral olfaction model.}
\end{center}
\end{table}
\noindent If we choose the characteristic frequency of the chiral odorant close to the typical vibrational frequency of an odorant, our results would be numerically consistent with the results of previous works~\cite{Bro2,Che}: for $\delta=10^{13}Hz$ and $\omega_{z}=10^{12}Hz$, we obtain the inelastic ET time for left- and right-handed enantiomers as $0.77~ns$ and $8.6~ns$, and the elastic tunneling time as $80~ns$. The ratio of inelastic rate to elastic rate against the tunneling frequency for the left- and right-handed odorant with a constant localization frequency is plotted in FIG.~\ref{Fig2}. It obviously shows that the inelastic to elastic ratio increases with the ratio of tunneling frequency to localization frequency, but with different rates for two enantiomers. Since we are primarily interested in the odorant-assisted inelastic tunneling, the desired regime would be $\delta>\omega_{z}$.
\noindent The ratio of the inelastic rate for the left-handed enantiomer to the inelastic rate for the right-handed enantiomer with the same localization frequency, as plotted in FIG.~\ref{Fig3}, shows a minimum at $\delta\approx\omega_{z}$. In the limit $\delta>\omega_{z}$ where inelastic ET is favored over elastic one, the ratio increases with the ratio of tunneling frequency to localization frequency. We suggest that this difference is the basis of chiral discrimination in olfactory system. 
\section{V. Conclusion}
A common reason to reject vibration-based theories of olfaction is the existence of enantiomers of chiral odorants with the same vibrational spectra, but different smells. Despite the same vibrational spectra, there is an energy difference between the ground states of two enantiomers resulted from t the chiral intermolecular interactions between the odorant and the receptor. These chiral interactions might be the origin of chiral discrimination in olfactory system. Motivated by this idea, we have developed a dynamical model to examine the chiral odorant-mediated electron transport in the context of a proposed olfactory process. We modeled the chiral odorant effectively as an asymmetric double-well potential. The introduced asymmetry, introduced as localization frequency, is considered as an overall measure of all chiral interactions. The resulting (in)elastic ET rates are presented in (\ref{25}) and (\ref{26}). Our results show that if we choose the parameters of our model close to the corresponding parameters of previous works on vibrational olfaction, we obtain consistent values for the inelastic rates, yet different for two enantiomers. More specifically, the ratio of tunneling frequency to localization frequency can be considered as the discrimination factor of our model. We showed that an increase in this factor intensifies the inelastic tunneling for both enantiomers (FIG.~\ref{Fig2}). At the region where inelastic tunneling is dominant, chiral discrimination increases with the discrimination factor (FIG.~\ref{Fig3}). It is an experimental fact that the majority of enantiomer pairs have very similar smells. In the model we proposed this is realized by the fact that majority of chiral odorants have low ratio of tunneling frequency to localization frequency, and therefore weak chiral discrimination by the olfactory system.

\end{document}